\documentclass[aps,superscriptaddress,nobibnotes,prb,twocolumn]{revtex4-2}
\usepackage{graphicx}
\usepackage{float}
\usepackage{bm}
\usepackage{color}
\usepackage[squaren]{SIunits}
\usepackage{amssymb}
\usepackage{amsmath}
\usepackage{epstopdf}
\usepackage{natbib}
\usepackage[colorlinks,linkcolor=blue,citecolor=blue,urlcolor=blue]{hyperref}
\usepackage{array}
\setcitestyle{open={[},close={]},citesep={,\!},numbers}

\begin{document}
	
\title{Fermiology with nodal structures in nonsymmorphic superconductor $\mathbf{LaNiGa_{\text{2}}}$: A de Haas-van Alphen study}
\author{Houpu Li}
\affiliation{Department of Physics, University of Science and Technology of China, Hefei 230026, China}

\author{Ye Yang}
\affiliation{Hefei National Research Center for Physical Sciences at the Microscale, University of Science and Technology of China, Hefei 230026, China}
\affiliation{School of Physics, Hefei University of Technology, Hefei, Anhui 230009, China}

\author{Mengzhu Shi}
\affiliation{Hefei National Research Center for Physical Sciences at the Microscale, University of Science and Technology of China, Hefei 230026, China}

\author{Yingcai Qian}
\affiliation{Department of Physics, University of Science and Technology of China, Hefei 230026, China}
\affiliation{High Magnetic Field Laboratory, Chinese Academy of Sciences, Anhui 230031, China}

\author{Senyang Pan}
\affiliation{Department of Physics, University of Science and Technology of China, Hefei 230026, China}
\affiliation{High Magnetic Field Laboratory, Chinese Academy of Sciences, Anhui 230031, China}

\author{Kaibao Fan}
\affiliation{Hefei National Research Center for Physical Sciences at the Microscale, University of Science and Technology of China, Hefei 230026, China}

\author{Nan Zhang}
\affiliation{Department of Physics, University of Science and Technology of China, Hefei 230026, China}

\author{Kaixin Tang}
\affiliation{Hefei National Research Center for Physical Sciences at the Microscale, University of Science and Technology of China, Hefei 230026, China}

\author{Hongyu Li}
\affiliation{Hefei National Research Center for Physical Sciences at the Microscale, University of Science and Technology of China, Hefei 230026, China}

\author{Zhiwei Wang}
\affiliation{Department of Physics, University of Science and Technology of China, Hefei 230026, China}

\author{Jinglei Zhang}
\affiliation{High Magnetic Field Laboratory, Chinese Academy of Sciences, Anhui 230031, China}

\author{Chuanying Xi}
\affiliation{High Magnetic Field Laboratory, Chinese Academy of Sciences, Anhui 230031, China}

\author{Ziji Xiang}
\altaffiliation{E-mail: zijixiang@ustc.edu.cn}
\affiliation{Hefei National Research Center for Physical Sciences at the Microscale, University of Science and Technology of China, Hefei 230026, China}
\affiliation{Hefei National Laboratory, University of Science and Technology of China, Hefei 230088, China}
	
\author{Xianhui Chen}
\altaffiliation{E-mail: chenxh@ustc.edu.cn}
\affiliation{Department of Physics, University of Science and Technology of China, Hefei 230026, China}
\affiliation{Hefei National Research Center for Physical Sciences at the Microscale, University of Science and Technology of China, Hefei 230026, China}
\affiliation{Hefei National Laboratory, University of Science and Technology of China, Hefei 230088, China}
	
\begin{abstract}
    Topological metals possess various types of symmetry-protected degenerate band crossings. When a topological metal becomes superconducting, the low-energy electronic excitations stemming from the band crossings located close to the Fermi level may contribute to highly unusual pairing symmetry and superconducting states. In this work, we study the electronic band structure of the time-reversal symmetry breaking superconductor LaNiGa$_2$ by means of quantum oscillation measurements. A comprehensive investigation combining angle-resolved high-field de Haas-van Alphen (dHvA) spectroscopy and first-principles calculations reveals the fermiology of LaNiGa$_2$ and verifies its nonsymmorphic $Cmcm$ lattice symmetry, which promises nodal band crossings pinned at the Fermi level with fourfold degeneracies. Moreover, such nodal structures, proposed to play a crucial role giving rise to the interorbital triplet pairing, are indeed captured by our dHvA analysis. Our results identify LaNiGa$_2$ as a prototypical topological crystalline superconductor and highlight the putative contribution of low-energy nodal quasiparticles to unconventional superconducting pairing.
		
\end{abstract}
	
\maketitle

\section{Introduction}
	
  While in all superconductors the superconducting transition marks a spontaneously breaking of the (global) $U(1)$ gauge symmetry \cite{BCS,AndersonGauge,AnnettSym}, additional symmetries, such as the lattice point-group symmetry, the time-reversal symmetry and the rotational symmetry, can be broken in the superconducting state of numerous superconductors \cite{UNSC}. The latter category represents the so-called ``unconventional superconductors" which have been an intense focus of research. Among them, superconductors with time-reversal symmetry breaking (TRSB) \cite{RecentTRSB} are particularly remarkable: their superconducting states implicitly host a spontaneous internal magnetic field, challenging the conventional wisdom that magnetic fields are always detrimental to superconductivity. This small spontaneous field can be detected by the zero-field muon-spin rotation ($\mu$SR) \cite{CaPtAs,mSR1,mSR2} or polar Kerr effect \cite{kerr1,kerr2} measurements, thus serves as the most prominent experimental evidence for TRSB superconductivity. Despite its unconventional nature, the realization of TRSB superconductivity does not rely on exotic pairing mechanisms that beyond the Bardeen-Cooper-Schrieffer (BCS) theory; one of the key ingredients is the interband interaction in multiband superconductors, which adds a degree of freedom to the superconducting wave function and helps to create TRSB effect in the pairing state \cite{ConvenTRSB,Tanaka}. When the interband interaction comes into play, the time-reversal symmetry could be broken in two different ways \cite{RecentTRSB}: (i) Considering the band (orbital) channel, the repulsive interband interaction can invoke an arbitrary phase difference (neither 0 nor $\pi$) between superconducting gaps on individual bands, giving rise to a complex, multicomponent superconducting order parameter that breaks the time-reversal symmetry \cite{ConvenTRSB,Threeband,LeggettMode}; (ii) alternatively, TRSB can occur in the spin channel, owing to the realization of a unique equal-spin (triplet) pairing state prompted by strong interband/interorbital pairing strength \cite{DaiLaFeAsOF,LaNiGa2TwoGap,Interband1,Interband2}.

  In most multiband superconductors, the interband/interorbital pairing potential is still too weak to produce TRSB superconductivity. Intriguingly, the case can be different in superconductors with topologically nontrivial band structures, wherein the symmetry-protected nodal band touchings (e.g., nodal points and nodal lines) associated with various topological invariants naturally guarantee band degeneracies \cite{Topology1,Topology2}; once these topological nodal structures present in the vicinity of the Fermi level ($E_F$), they can significantly promote the interband/interorbital pairing \cite{Interband1,Interband2} because now the direct interband transitions can persist all the way to zero energy \cite{LaNiGa2ARPES1,QuanTopology}, explicitly invalidating the single-band picture. Recent studies has revealed the TRSB superconductor $\mathrm{LaNiGa_{\text{2}}}$ to be a promising candidate of ``topological crystalline superconductors" in which the Dirac-type (fourfold degenerate) band crossings protected by nonsymmorphic lattice symmetry are suggested to be essential for the occurrence of unconventional superconducting state comprising interband triplet pairs \cite{LaNiGa2ARPES1,QuanTopology,LaNiGa2ARPES2}, thus identify this material as an ideal platform for investigating the intertwinement between exotic pairing and nontrivial band topology.

\begin{figure*}[t]%"[]"中为位置参数，四个参数tbph依次是置顶、置底、浮动、当前位置，，选用的参数优先顺序为h-t-b-p
\centering
\includegraphics[width=2.0\columnwidth]{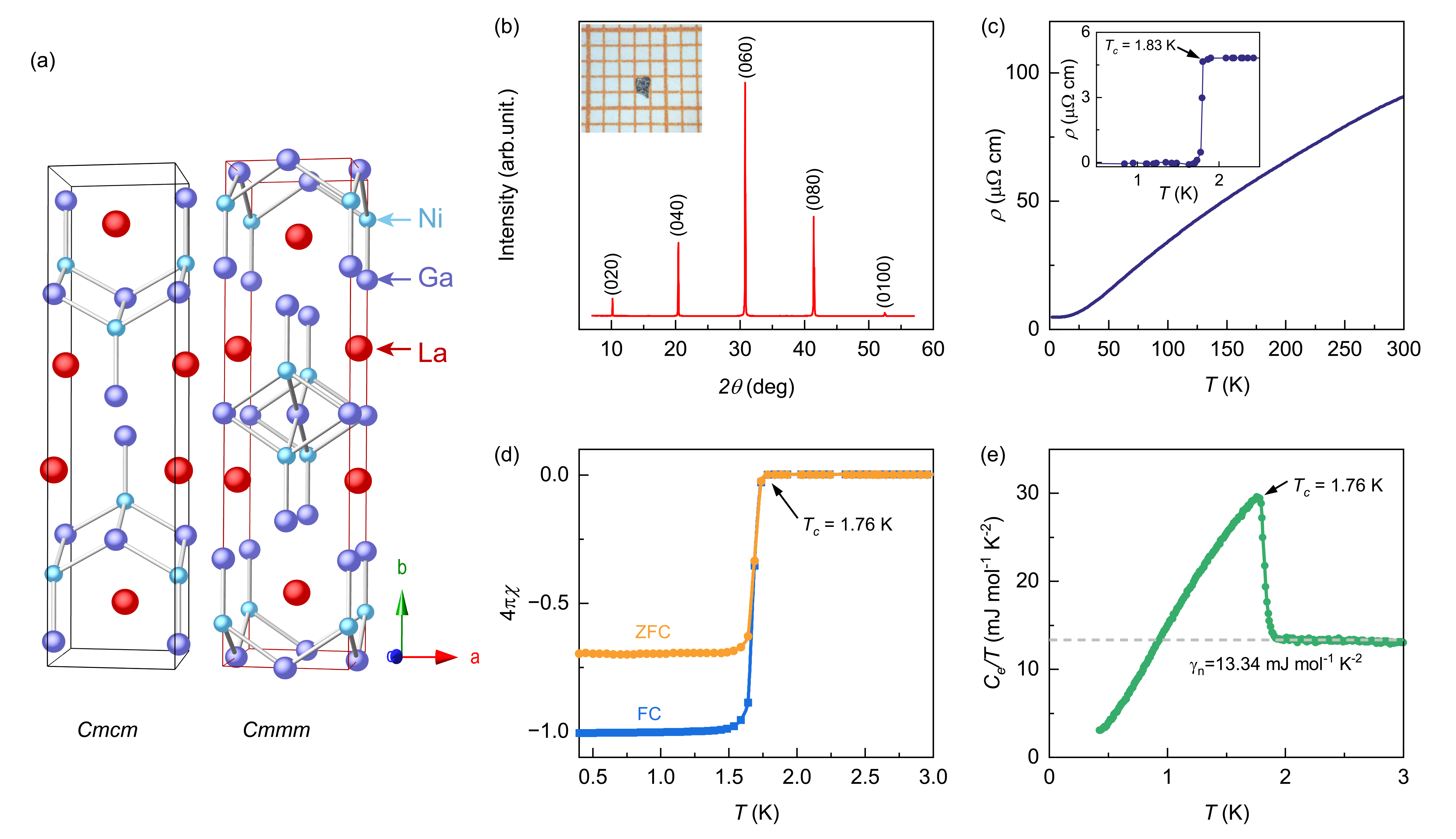}%"scale"后的数字为图形的宽度，也可用"width=1.0\columnwidth"定义
\caption{(Color online) Structural and superconducting properties of $\mathrm{LaNiGa_{\text{2}}}$ single crystals. (a) Nonsymmorphic $Cmcm$ (left) and symmorphic $Cmmm$ (right) unit cells of $\mathrm{LaNiGa_{\text{2}}}$. The bonds between Ni (cyan) and Ga (navy) are depicted to highlight the difference. (b) The signal crystal X-ray diffraction (XRD) pattern of a $\mathrm{LaNiGa_{\text{2}}}$ sample, showing the ($0k0$) series of Bragg reflections. Inset: an optical image of single crystalline LaNiGa$_2$. (c) Temperature dependence of electrical resistivity $\rho$, measured between 2 and 300\,K with current applied in the (010) plane. Inset shows the low-temperature resistivity data, highlighting the resistive superconducting transition at 1.83\,K. (d) Zero-field cooled (ZFC, blue) and field cooled (FC, yellow) magnetic susceptibility, $\chi = M/H$, where $M$ is the magnetization (per unit volume) and $H$ the applied magnetic field. Data were measured with $\mu_0H$ = 1\,mT. (e) Zero-field temperature dependence of the electronic specific heat $C_e/T$; the lattice contribution of  specific heat was subtracted from the raw data (see Supporting Information Fig.\,S2 for raw data). The dashed line denotes the extraction of Sommerfeld coefficient $\gamma_n$ for $\mathrm{LaNiGa_{\text{2}}}$.} %图题
\label{fig1}%{}中"fig:example1"为图名，引用时用\ref{fig:example1}
\end{figure*}

  $\mathrm{LaNiGa_{\text{2}}}$ is a nonmagnetic compound \cite{LaNiGa2NMR} that becomes superconducting below $T_c \approx$ 2\,K \cite{LaNiGa2SC}. Although its superconductivity can be delineated by a nodeless two-gap model, ostensibly resembling a (multiband) weakly-coupled isotropic BCS superconductor \cite{LaNiGa2TwoGap,LaNiGa2SC,SundarMuon}, zero-field $\mu$SR experiment has revealed spontaneous magnetic fields below $T_c$ \cite{LaNiGa2Nonunitary}, suggesting a highly unusual superconducting state with TRSB. Such a fully-gapped TRSB pairing state has been understood with a proposal of the internally antisymmetric nonunitary triplet (INT) pairing \cite{LaNiGa2TwoGap,LaNiGa2SCTheory}: in the INT scenario, the equal-spin paring occurs between electrons on the same site but on different orbitals, giving rise to a multicomponent superconducting order parameter consisting of nonequal $\uparrow\uparrow$ and $\downarrow\downarrow$ pairing gaps; the spontaneous internal field probed by zero-field $\mu$SR is thus attributed to the migration of spin-polarized Cooper pairs between the minority and majority condensates \cite{LaNiGa2SCTheory}. Later studies on single crystalline samples indicate that the lattice structure of $\mathrm{LaNiGa_{\text{2}}}$ has nonsymmorphic symmetry \cite{LaNiGa2ARPES1} [space group $Cmcm$, No.\,63, instead of the symmorphic space group $Cmmm$/No.\,65 determined from polycrystalline samples \cite{LaNiGa2SC,LaNiGa2Cmmm}, see sketches in Fig.\,\ref{fig1}(a)]. The nonsymmorphic structure ensures Dirac line nodes (supposedly gapped by spin-orbit coupling (SOC)) and Dirac points (surviving the SOC effect) that are pinned at $E_F$ \cite{LaNiGa2ARPES1}; these topological band degeneracies lead to strong interband coupling substantial for realizing the INT pairing \cite{QuanTopology,SundarMuon}. Nevertheless, it still requires more decisive evidence to pinpoint the nonsymmorphic symmetry and topological metal nature of LaNiGa$_2$, since the difference between $Cmmm$ and $Cmcm$ space groups is rather weak for an X-ray diffraction (XRD) determination \cite{Structure1,Structure2}, whilst in angle-resolved photoemission spectroscopy (ARPES) experiments \cite{LaNiGa2ARPES1} it appears to be challenging to identify the linear band crossings at $E_F$. Here, thanks to the successful growth of high-quality $\mathrm{LaNiGa_{\text{2}}}$ single crystals, we observe the dHvA effect, i.e., quantum oscillations in magnetization, for the first time in this compound using torque magnetometry technique under intense magnetic fields ($H$). The clearly resolved dHvA effect provides a good opportunity to look into the $k$-space electronic structure with high precision. Combining dHvA analysis with first-principles calculations, we construct the detailed fermiology of $\mathrm{LaNiGa_{\text{2}}}$ and confirm the nonsymmorphic $Cmcm$ symmetry, whose topological features in band structure are unequivocally reflected in the dHvA spectra.

\section{Materials and methods}
	
  Single crystals of $\mathrm{LaNiGa_{\text{2}}}$ were grown using the Ga deficient self-flux technique \cite{LaNiGa2ARPES1}. High-purity ingots of elements La (99.99\%), Ni (99.999\%) and Ga (99.999\%) were mixed at an atomic ratio of $0.33:0.33:0.34$; precursor ingots were first synthesized by arc melting the elements several times in a water-cooled argon arc furnace (total mass loss during this process was determined to be less than 1\%). The ingots were then loaded into a boron-nitride (BN) crucible. We did not use alumina crucible because it introduces excess aluminum into the crystalline lattice which substitutes gallium and renders the sample nonstoichiometric (see the energy-dispersive X-ray spectrum data in Supporting Information Fig.\,S1 for details). The BN crucible was sealed in an evacuated quartz tube and heated at $1150$ $\mathrm{^\circ C}$ for $10$ h, followed by slow cooling down to $800$ $\mathrm{^\circ C}$ over $100$ h. The excess flux was decanted using a centrifuge. The obtained single crystals were mostly in a plate-like shape with typical dimensions of $1 \times 1 \times 0.2$ $\mathrm{mm^3}$.

\begin{figure*}[t]%"[]"中为位置参数，四个参数tbph依次是置顶、置底、浮动、当前位置，，选用的参数优先顺序为h-t-b-p
\centering
\includegraphics[width=2.0\columnwidth]{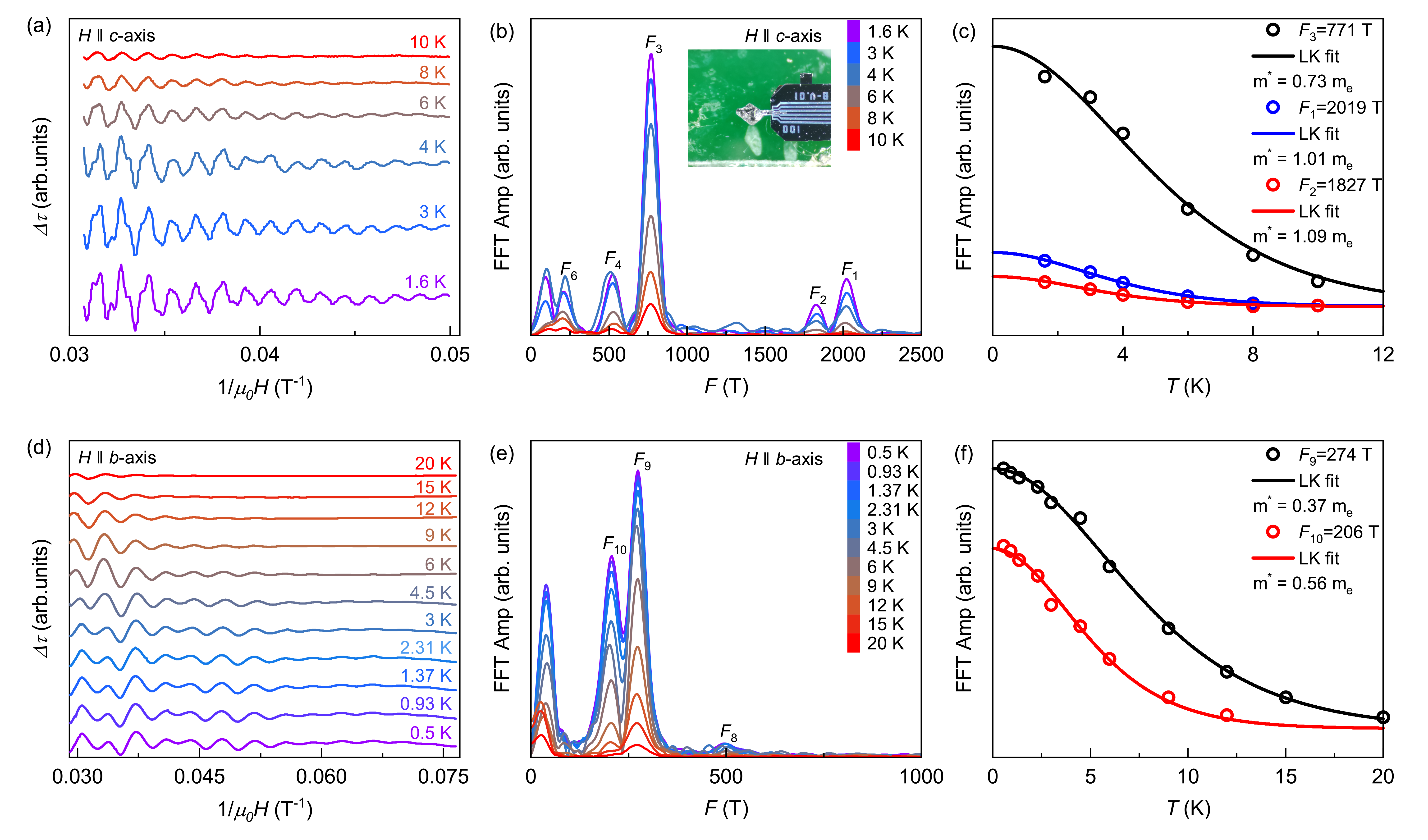}%"scale"后的数字为图形的宽度，也可用"width=1.0\columnwidth"定义
\caption{(Color online) Temperature dependent dHvA oscillations measured with field along two crystalline axes ($b$ and $c$). (a) Oscillatory magnetic torque $\Delta\tau$ measured at varying temperatures with $H \parallel c$, plotted as a function of inverse magnetic field. To obtain $\Delta\tau$, a polynomial background was subtracted from the signal voltage of the piezoresistive cantilever (see Supporting Information Fig.\,S3 for raw data). (b) The fast Fourier transform (FFT) spectra corresponding to the dHvA oscillations shown in (a). The field range for FFT window is 20-32.5\,T. Inset shows a $\mathrm{LaNiGa_{\text{2}}}$ single crystal mounted on the SCL piezoresistive cantilever. (c) Lifshitz-Kosevich (LK) fits of the $T$ dependence of dHvA oscillation amplitudes (defined as the FFT peak heights) for selected branches. (d)-(f) Same as (a)-(c) but data were measured with $H \parallel b$. The field range for FFT window in (e) is 13-33\,T.} %图题
\label{fig2}%{}中"fig:example1"为图名，引用时用\ref{fig:example1}
\end{figure*}

  The crystallographic orientation of $\mathrm{LaNiGa_{\text{2}}}$ single crystals was examined by XRD measurements performed on a Bruker D8 diffractometer (with Cu $K_{\rm \alpha1}$ radiation). The in-plane orientations of samples were determined using a Huber Laue 9920 single-crystal diffraction detector. Electrical transport measurements were performed using a standard four-probe ac method in a Quantum Design DynaCool-$14$ T Physical Property Measurement System (PPMS); the electrical contacts were made by attaching Pt wires to the sample using silver paste (Dupont 4922N). Specific heat measurements were also performed in PPMS utilizing a $^3$He insert. Magnetic susceptibility in the superconducting state was measured down to $0.4$ K in a Magnetic Property Measurement System (MPMS-XL SQUID, Quantum Design) equipped with an $^3$He refrigerator. The dHvA studies were conducted based on magnetic torque data probed using a SCL piezoresistive self-sensing cantilever placed on a rotation stage. A photograph showing a $\mathrm{LaNiGa_{\text{2}}}$ crystal attached to the SCL cantilever is presented in the inset of Fig.\,\ref{fig2}(b). Angle-resolved torque magnetometry experiments were performed in a $^3$He cryostat placed in a water-cooled Bitter magnet supplying dc magnetic fields up to 33\,T at the Chinese High Magnetic Field Laboratory (CHHML) in Hefei.

  Density Functional Theory (DFT) calculations of band structure were performed by employing the Vienna ab initio simulation package (VASP) \cite{DFT}. The Perdew-Burke-Ernzerh (PBE) functional \cite{PBE} described the exchange correlation energy. The cutoff energy of the plane-wave basis was set to $450$ eV. The Brillion Zone (BZ) was sampled using the $12 \times 12 \times 12$ Monkhorst-Pack grids for the self-consistent calculation. To investigate the Fermi surface, we constructed the Wannier tight-binding Hamiltonian based on the projected Ni 3d and Ga 4p orbitals as implemented in the WANNIER90 package \cite{Wannier}. The angular dependence of Fermi surface extremal cross-sectional areas was computed via the SKEAF program \cite{SKEAF}.

\section{Results and discussion}
	
  In Fig.\,\ref{fig1}(a) we compare the nonsymmorphic $Cmcm$ and symmorphic $Cmmm$ crystal structure of $\mathrm{LaNiGa_{\text{2}}}$. Both structures are orthorhombic. The lattice parameters are $a$ = 4.273\,\AA, $b$ = 17.412\,\AA, $c$ = 4.268\,\AA\, for the $Cmcm$ unit cell and $a$ = 4.278\,\AA, $b$ = 17.436\,\AA, $c$ = 4.271\,\AA\, for the $Cmmm$ unit cell \cite{LaNiGa2SC,LaNiGa2ARPES1}; owing to the small difference in lattice parameters, it is difficult to distinguish the two space groups from single-crystal XRD patterns shown in Fig.\,\ref{fig1}(b), which reveals the (010) orientation of our $\mathrm{LaNiGa_{\text{2}}}$ crystals (i.e., the $b$-axis is perpendicular to the top surface). The temperature ($T$) dependence of resistivity displayed in Fig.\,\ref{fig1}(c) implies lower residual resistivity (4.8\,$\mu\Omega$ cm) and larger residual resistivity ratio $\rho$(300\,K)/$\rho$(2\,K) ($\sim$ 20) compared with single crystal data presented in Ref.\,\cite{LaNiGa2ARPES1}, suggesting higher crystal quality for our samples. Low-$T$ resistivity data reveals a sharp superconducting transition occurring at $T_c$ = 1.83\,K [inset of Fig.\,\ref{fig1}(c), see also Fig.\,S1 in Supporting Information], whereas the onset of bulk superconductivity determined from magnetic susceptibility [Fig.\,\ref{fig1}(d)] and specific heat [Fig.\,\ref{fig1}(e)] measurements is approximately 1.76\,K. The value of $T_c$ in our samples is similar to that in high-quality polycrystalline LaNiGa$_2$ prepared by long-time annealing \cite{LaNiGa2TwoGap}, but appears to be lower than the previously reported single crystal results \cite{LaNiGa2ARPES1,SundarMuon} (typically between 2.0 and 2.1\,K). We note that the higher $T_c$ is probably related to the Al doping introduced by alumina crucible during the growth process, which is absent in our samples (see Supplementary Information Fig.\,S1 for details). Nevertheless, the zero-field cooled (ZFC) magnetic susceptibility [Fig.\,\ref{fig1}(d)] indicates full magnetic shielding; the small separation between the ZFC and filed-cooled (FC) curves points towards a flux pinning effect much weaker than that in polycrystalline sample \cite{LaNiGa2SC} due to the reduced number of pinning centers. The specific heat jump at $T_c$, $\Delta C_e/\gamma_nT_c$ = 1.32, and the normal-state Sommerfeld coefficient $\gamma_n$ = 13.34 mJ/mol K$^2$ [Fig.\,\ref{fig1}(e)] are in good agreement with earlier results \cite{LaNiGa2SC,LaNiGa2ARPES1,LaNiGa2TwoGap}. More details of specific heat analysis are presented in Supporting Information Fig.\,S2.

  To further explore the electronic structure of $\mathrm{LaNiGa_{\text{2}}}$, we conduct angle-resolved quantum oscillation study which is a powerful tool for mapping the Fermi surface (FS) morphology in metals \cite{Shoenberg}. The dHvA effect was measured by torque magnetometry using an SCL piezoresistive sensor as mentioned above; the orientation of magnetic field with respect to the crystal axes was adjusted by a manually-controlled horizontal rotator. Figures \ref{fig2}(a) and \ref{fig2}(d) show profiles of the oscillatory magnetic torque $\Delta\tau$ measured at various temperatures with $H \parallel c$ (in-plane) and $H \parallel b$ (out-of-plane), respectively. Note that the nonoscillatory voltage signal detected by the SCL cantilever in an $H$-field contains a spurious ``magnetoresistance" background arising from the Wheatstone bridge circuit inside the sensor, thus does not accurately reflect the intrinsic magnetic torque of the sample. These oscillatory components incorporate all the crucial information for a dHvA study. Within the framework of three-dimensional Lifshitz-Kosevich (LK) theory \cite{Shoenberg}:
\begin{align}
\Delta\tau = & A_{0}B^{3/2}\sum_{i}\frac{dF_{i}}{d\theta}\Big|\frac{\partial^{2}S_{i,k}}{\partial k_{\parallel}^{2}}\Big|^{-1/2}\sum_{p=1}^{\infty}p^{-3/2}R_{T}R_{D}R_{S}\notag\\&sin[2\pi p(\frac{F_{i}}{B}+\frac{\Phi_{i}}{2\pi})\pm\frac{\pi}{4}],
\label{eq1}
\end{align}

\begin{figure*}[t]%"[]"中为位置参数，四个参数tbph依次是置顶、置底、浮动、当前位置，，选用的参数优先顺序为h-t-b-p
\centering
\includegraphics[width=2.0\columnwidth]{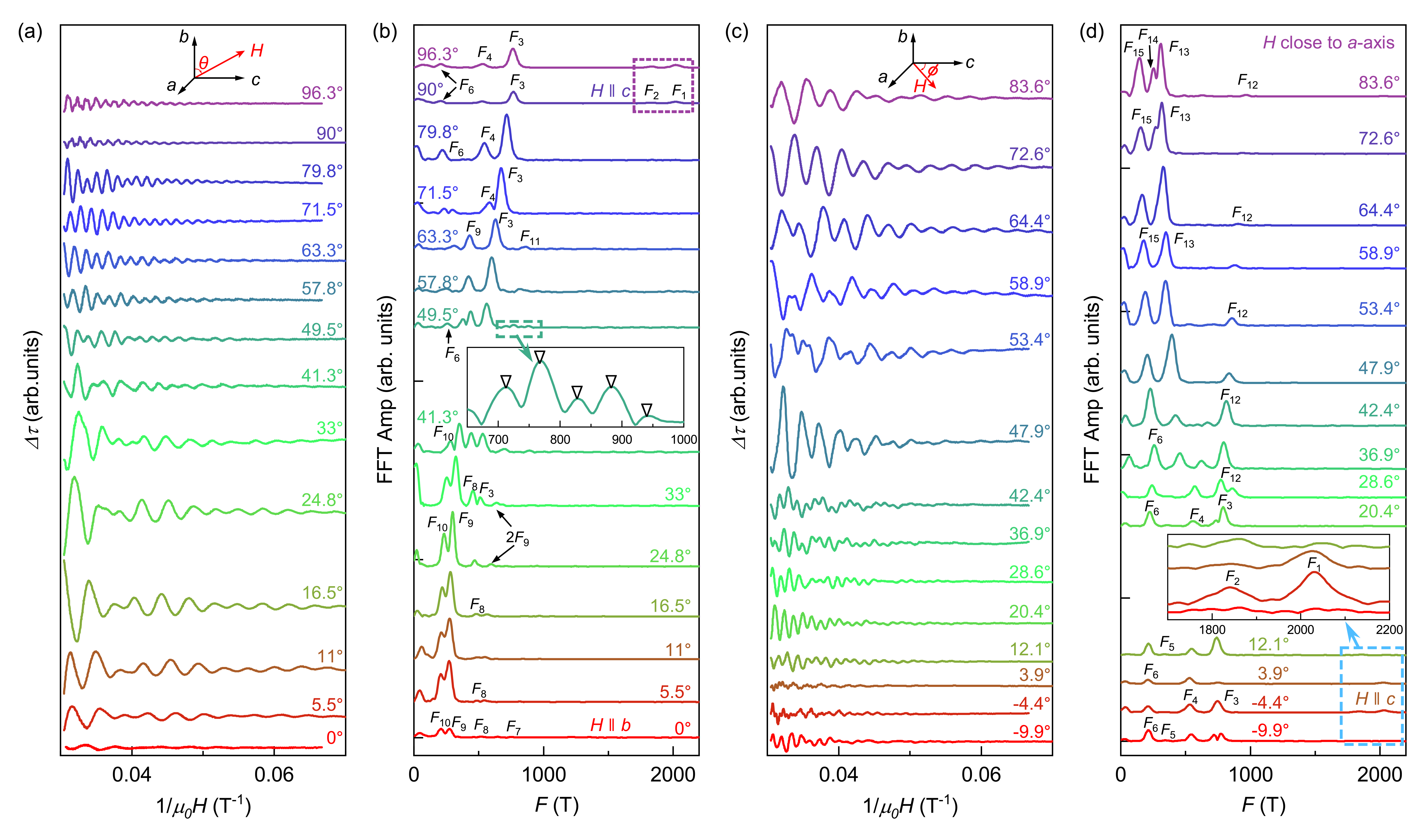}%"scale"后的数字为图形的宽度，也可用"width=1.0\columnwidth"定义
\caption{(Color online) Angle-resolved dHvA effect measured at $T$ = 1.6\,K. (a) $\Delta\tau$(1/$\mu_0H$) taken at varying angles $\theta$, which is defined as the angle between the field vector $H$ and the $b$-axis. During the measurement, $H$ was rotated in the $bc$-plane as depicted by the inset sketch. (b) The FFT spectra of the dHvA data shown in (a). (c) $\Delta\tau$(1/$\mu_0H$) taken at varying angles $\phi$ defined as the angle between $H$ and the $c$-axis. During the measurement, $H$ was rotated in the $ac$-plane as depicted by the inset sketch. (d) The FFT spectra of the dHvA data shown in (c). The FFT peaks in (b) and (d) are labeled with indices $F_1$-$F_{15}$. Insets in (b) and (d) give expanded views of some relatively weak FFT peaks. The short-dashed boxes highlight the high-$F$ components, $F_1$ and $F_2$. Dashed box in (b) marks the frequency range where the putative magnetic breakdown peaks show up. } % 图题
\label{fig3}%{}中"fig:example1"为图名，引用时用\ref{fig:example1}
\end{figure*}

  here $A_0$ is a constant; the outer sum and the inner sum run over band indices $i$ and oscillation harmonics $p$, respectively; $S_i$ is the extremal cross-sectional of the $i$-th Fermi surface; $k_{\parallel}$ is the momentum component parallel to $H$; $R_T$, $R_D$ and $R_S$ are the thermal, Dingle and spin-splitting damping factors, respectively; $\Phi_i$ is the total phase of quantum oscillation for the $i$-th band \cite{FeGe}. In this study we take the magnetic induction $B\approx\mu_0H$ considering the small magnetization $M$ of $\mathrm{LaNiGa_{\text{2}}}$. The Onsager relation directly links the dHvA frequency $F_i$ to $S_i$: $F_i=\displaystyle\frac{\hbar}{2\pi e}S_i$. To resolve $F_i$, we perform the fast Fourier transform (FFT) on the $\Delta\tau$(1/$\mu_0H$) data and the results are displayed in Figs.\,\ref{fig2}(b) and \ref{fig2}(e). With $H \parallel c$, two high-frequency ($F$) dHvA branches, i.e., $F_1$ = 2019\,T and  $F_2$ =1827\,T, are resolved [Fig.\,\ref{fig2}(b)]; these frequencies correspond to cyclotron orbits lying in the nodal plane $k_z = \pi/c$ \cite{LaNiGa2ARPES1,QuanTopology} and their presence strongly supports the nonsymmorphic $Cmcm$ symmetry, as we shall show below. Three more branches with frequencies lower than 1\,kT are observed for this field orientation, labeled as $F_3$, $F_4$ and $F_6$ in Fig.\,\ref{fig2}(b). For the $H \parallel b$ configuration, only dHvA branches with $F \lesssim$ 500\,T are unambiguously resolvable, with the two strongest FFT peaks $F_9$ = 274\,T and $F_{10}$ =206\,T corresponding to small orbits [Fig.\,\ref{fig2}(e)]. We mention that the FFT peaks below 100\,T in Figs.\,\ref{fig2}(b) and \ref{fig2}(e) vary considerably with the selection of the FFT window and are likely to be artifacts due to incomplete background subtraction. From the $T$ dependence of FFT amplitudes, we can obtain the information of effective cyclotron mass of electrons by fitting the data to the standard LK formalism \cite{Shoenberg}, in which the thermal damping factor $R_T$ is given by:
\begin{align}
R_T(T) \propto \frac{\frac{\alpha m^*T}{m_e\mu_0H}}{\sinh(\frac{\alpha m^*T}{m_e\mu_0H})},
\label{eq2}
\end{align}
  where $m^*$ and $m_e$ are the cyclotron mass and the free electron mass, respectively, $\alpha \equiv$ 14.69\,T/K. The value of 1/$H$ in Eq.\,\ref{eq2} is set to the averaged inverse magnetic field in the FFT field range. The fits shown in Figs.\,\ref{fig2}(c) and \ref{fig2}(e) yields $m^*/m_e$ = 1.01 ($F_1$), 1.09 ($F_2$), 0.73 ($F_3$), 0.37 ($F_9$) and 0.56($F_{10}$). The moderate values of $m^*$ corroborate the absence of appreciable electron correlation in $\mathrm{LaNiGa_{\text{2}}}$ \cite{LaNiGa2NMR}.

 \begin{figure*}[t]%"[]"中为位置参数，四个$m_e$参数tbph依次是置顶、置底、浮动、当前位置，，选用的参数优先顺序为h-t-b-p
\centering
\includegraphics[width=2.0\columnwidth]{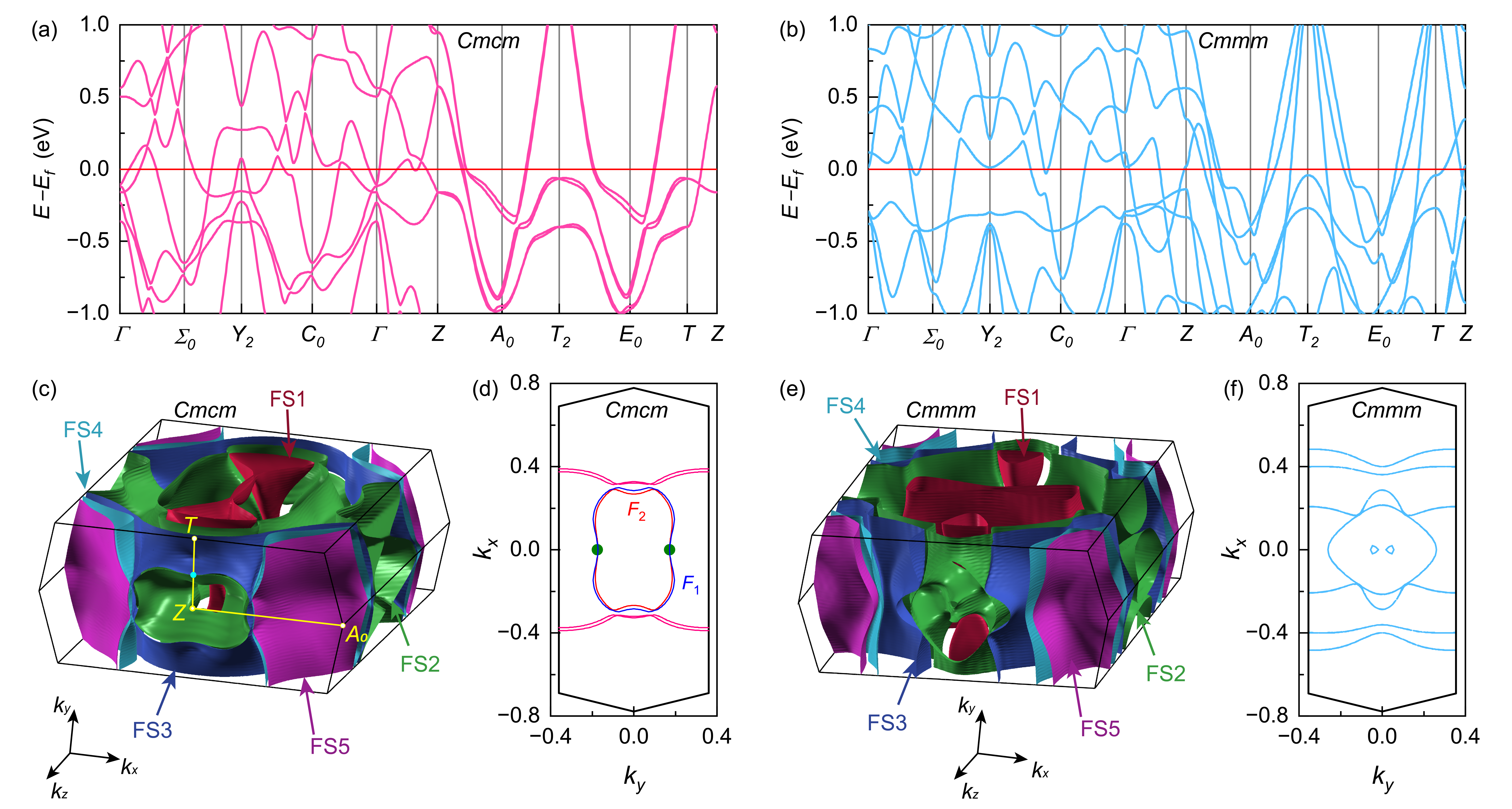}%"scale"后的数字为图形的宽度，也可用"width=1.0\columnwidth"定义
\caption{(Color online) DFT-calculated band structure and Fermi surface for the $Cmcm$ and $Cmmm$ space groups. (a), (b) Band structure of $\mathrm{LaNiGa_{\text{2}}}$ with the nonsymmorphic $Cmcm$ and symmorphic $Cmmm$ symmetries, respectively, obtained from DFT calculations with SOC. For the position of high symmetry points within the first Brillouin zone, see Supporting Information Fig.\,S4. (c), (e) Fermi surface structure of $\mathrm{LaNiGa_{\text{2}}}$ for $Cmcm$ and $Cmmm$ structures, respectively. Both have five Fermi surfaces, labeled as FS1-5. (d), (f) Fermi surface contours on the $k_z = \pi/c$ plane for $Cmcm$ and $Cmmm$ structures, respectively. Green dots in (d) represent the Dirac points at which the fourfold degeneracy of electron states is topologically protected even with SOC (see the discussion in the main text).} % 图题
\label{fig4}%{}中"fig:example1"为图名，引用时用\ref{fig:example1}
\end{figure*}

  The angle-resolved dHvA investigation was performed by measuring magnetic torque on two samples with different configuration: (i) the $H$-vector is rotated in the $bc$ planes, with an inclination angle $\theta$ with respect to the $b$-axis; (ii) $H$ rotating in the $ac$ plane from $a$ to $c$ axis with a tilt angle $\phi$ [see the inset sketches in Figs.\,\ref{fig3}(a) and \ref{fig3}(c), respectively]. The oscillatory torque curves and their corresponding FFT spectra are summarized in Figs.\,\ref{fig3}(b) and \ref{fig3}(d). For both configurations, we find that the high-$F$ branches $F_1$ and $F_2$ appear only when $H$ is tilted away from the $c$-axis by less than 12$^\circ$ (i.e., $\theta$ close to 90$^\circ$ and $\phi$ close to 0$^\circ$), see the short-dashed boxes in Figs.\,\ref{fig3}(b) and \ref{fig3}(d). The low-$F$ dHvA components with $F <$ 1\,kT are weakly angle dependent, implying that they originate from three-dimensional small FS pockets. Moreover, in measurement configuration (i) we observe five additional FFT peaks [$F$ = 710, 768, 829, 882 and 942\,T, see the dashed box and inset in Fig.\,\ref{fig3}(b)] that suddenly emerge at $\theta$ = 49.5$^\circ$ near the original frequency $F_{11}$ and quickly become unresolvable at higher $\theta$. The peculiar $\theta$ dependence of these peaks and their almost even frequency spacing ($\Delta F \simeq$ 60\,T) strongly imply that they stem from magnetic breakdown (MB) orbits: when $H$ exceeds a threshold field, quasiparticle tunneling occurs between adjacent FSs, leading to a plethora of additional dHvA frequencies that are usually scaled with a linear combination of areas of semiclassical orbits \cite{Shoenberg,breakdown}. Nonetheless, locating the small orbit with $F \simeq$ 60\,T ($S \simeq$ 0.57 nm$^{-2}$) based on DFT calculation is challenging and identification of the MB orbits is beyond the scope of the present work.

\begin{figure*}[t]%"[]"中为位置参数，四个参数tbph依次是置顶、置底、浮动、当前位置，，选用的参数优先顺序为h-t-b-p
\centering
\includegraphics[width=2.0\columnwidth]{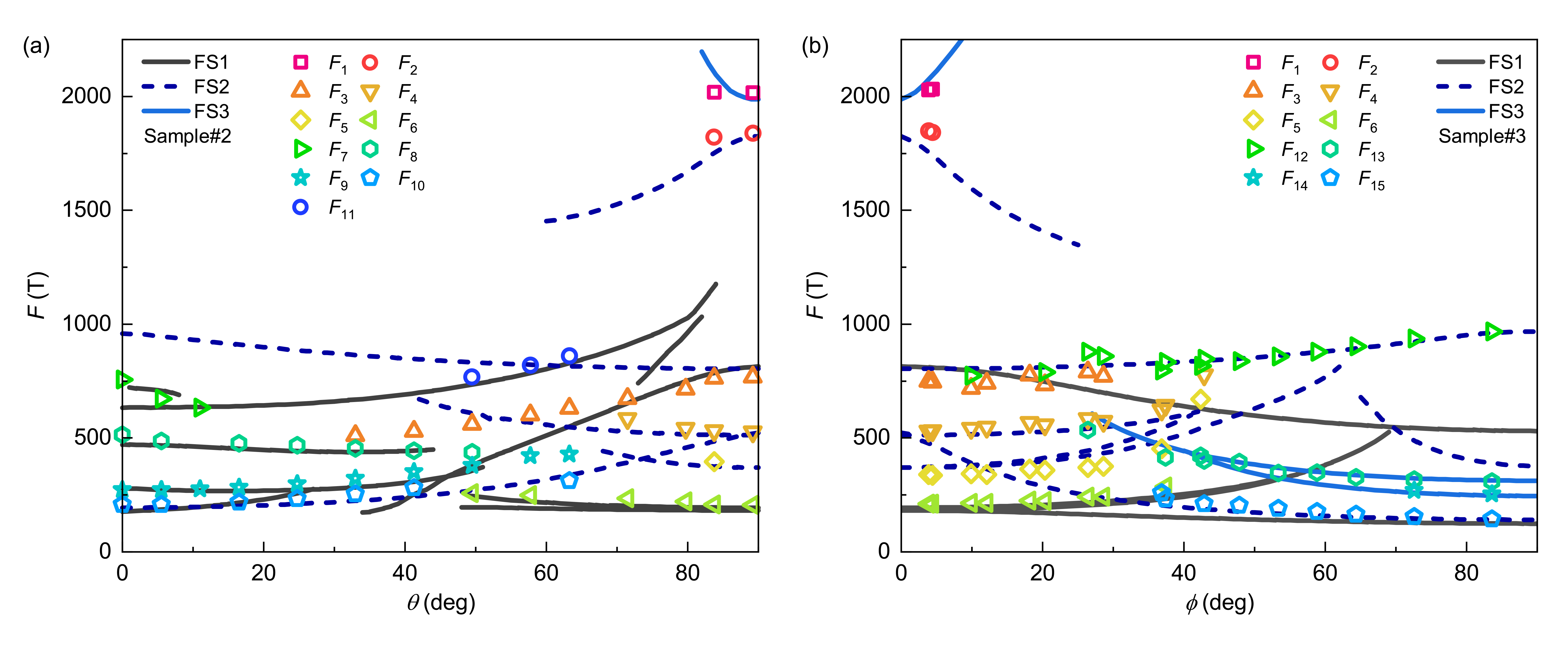}%"scale"后的数字为图形的宽度，也可用"width=1.0\columnwidth"定义
\caption{(Color online) Comparison between theoretically predicted $\theta$-dependent [(a)] and $\phi$-dependent [(b)] dHvA frequencies and the experimental data. The dHvA branches are calculated based on band structure for the $Cmcm$ symmetry [Fig.\,\ref{fig4}(a)] with $E_F$ shifted up by 50\,meV. Black solid lines, navy dashed lines and blue solid lines represent dHvA branches originating from extremal orbits on FS1, FS2 and FS3, respectively. The hollow symbols in (a) and (b) are dHvA frequencies extracted from FFT spectra presented in Figs.\,\ref{fig3}(b) and \ref{fig3}(d) [note that only the fundamental frequencies are displayed], respectively.} % 图题
\label{fig5}%{}中"fig:example1"为图名，引用时用\ref{fig:example1}
\end{figure*}

  Our results of DFT-calculated band structures for the nonsymmorphic $Cmcm$ and symmorphic $Cmmm$ space groups are displayed in Figs.\,\ref{fig4}(a) and \ref{fig4}(b), respectively; the calculations were performed based on reported structural information in previous works (see Supporting Information Table S1 and Table S2) \cite{LaNiGa2ARPES1,QuanTopology,SinghDFT} with spin-orbital coupling (SOC) taken into consideration. For both structures, there are five bands crossing $E_F$, giving rise to five sets of FSs \cite{SinghDFT,HaseLDA} as shown in Figs.\,\ref{fig4}(c) and \ref{fig4}(e), wherein FS1-FS5 are distinguished by colors. The most prominent difference between the two structures is on the $k_z = \pi/c$ plane, as manifested by the contours of FSs piercing through this plane depicted in Figs.\,\ref{fig4}(d) and \ref{fig4}(f). In the absence of SOC, the nonsymmorphic symmetry operations characterizing the $Cmcm$ structure (including a $c$ glide plane perpendicular to the $b$-axis and a $2_1$ screw axis along the $c$-axis \cite{LaNiGa2ARPES1,QuanTopology}) combining with TRS render $k_z = \pi/c$ a nodal surface, on which the bands must have fourfold degeneracy; in consequence, Dirac-type nodal lines and nodal loops (which are pinned at $E_F$) formed by crossing FS pairs are promised on this surface as all the FSs in $\mathrm{LaNiGa_{\text{2}}}$ are doubly degenerate. With SOC included, the fourfold band degeneracy on the nodal surface is lifted and most band crossings are erased by a gap of a few to 40\,meV (see Supporting Information Fig.\,S5 for details); nonetheless, two Dirac points at $k = (0, \pm0.48\frac{2\pi}{b}, \frac{\pi}{c})$ arising from touchings between FS2 and FS3 survive the SOC [green dots in Fig.\,\ref{fig4}(d)], thanks to the topological protection associated with the mirror reflection $M_x$ along $Z$-$T$ high symmetry line \cite{LaNiGa2ARPES1,Nodal_surface}. As a result, the nodal loop on the $k_z = \pi/c$ in the absence of SOC split into two loops crossing at these Dirac points, giving rise to two closed quasiparticle cyclotron orbits with similar areas [Fig.\,\ref{fig4}(d)]. In contrast, the band structure for the $Cmmm$ space group is completely free from such features: as shown in Fig.\,\ref{fig4}(f), there is only one large closed orbit on the $k_z = \pi/c$ plane.

  These unique features discriminating lattice symmetries and band topologies are relatively elusive in ARPES measurements because the (001) plane is not a natural cleaved surface for $\mathrm{LaNiGa_{\text{2}}}$ \cite{LaNiGa2ARPES1}. Fortunately, our angle-resolved dHvA experiments can capture them without ambiguity. In Figs.\,\ref{fig5}(a) and \ref{fig5}(b), the dHvA frequencies extracted from spectra presented in Figs.\,\ref{fig3}(b) and \ref{fig3}(d) are overlayed onto the theoretical dHvA branches predicted for the $Cmcm$ symmetry [corresponding to the DFT-calculated FSs shown in Fig.\,\ref{fig4}(c)], for measurement configurations (i) and (ii), respectively. Here, the position of $E_F$ in calculations was shifted upward by 50\,meV compared with the result shown in Fig.\,\ref{fig4}(a) to achieve better agreements with experimental data. We firstly focus on the origin of branches $F_1$ (2019\,T) and $F_2$ (1827\,T), the two high-$F$ components emerging when $H$ is aligned close to the crystalline $c$-axis (that is, almost perpendicular to the nodal surface mentioned above). As demonstrated by the consistency between calculations and measured results in Figs.\,\ref{fig5}(a) and \ref{fig5}(b), these two orbits can be convincingly assigned to the two FS loops on the $k_z = \pi/c$ plane split by SOC [Fig.\,\ref{fig4}(d)]. In contrast, orbits giving rise to these high-$F$ components are totally missing in the FSs of the $Cmmm$ structure with $H \parallel c$ (Supporting Information Fig.\,S6). Aside from $F_1$ and $F_2$, the low-$F$ dHvA frequencies are also satisfyingly reproduced by the model of DFT-calculated FSs based on the nonsymmorphic $Cmcm$ symmetry, as verified by the correspondence between data points and theoretical curves in Fig.\,\ref{fig5}. These small orbits reside on FS1, FS2 and FS3, whereas the orbits on the large pockets FS4 and FS5 (converting to $F >$ 5000\,T) are unobserved in our dHvA study; a detailed assignment of the dHvA frequencies to extremal FS orbits is presented in Supporting Information Fig.\,S4. On the other hand, the calculation referring to the $Cmmm$ symmetry failed to consistently track the angle dependence of experimental dHvA frequencies (Supporting Information Fig.\,S6).

  It is also worth noting that we did not resolve MB orbits stemming from orbits $F_1$ and $F_2$, which may reflect the unique feature of topological nodal structures that presumably link these two loops. According to the Chambers formula \cite{Chambers,Quasi-symmetry}, MB generally occurs above a threshold field $\mu_0 H_0 = (\pi/4\hbar e)(k_g^2 / v_{\parallel}v_{\perp})$ with a tunneling probability $P = \exp{(-H_0/H)}$ \cite{Shoenberg,Tunneling}, where $k_g$ is the energy gap for tunneling in the $k$-space, $v_{\parallel}$ and $v_{\perp}$ stand for the Fermi velocities along two in-plane directions perpendicular to the magnetic field at the breakdown junction. With the application of $H \parallel c$, both the TRS and the mirror symmetry $M_x$ are broken; consequently, the Dirac nodes on the $Z-T$ path should be gapped. Nevertheless, such symmetry-breaking effect results in a very tiny Dirac gap (less than 1\,meV even at 30\,T, see Supporting Information Fig.\,S7 for details) that yields a vanishingly small breakdown field $\mu_0 H_0 \lesssim$ 0.19\,T (Supporting Information Fig.\,S7). Hence, MB is expected to persist down to lowest fields in dHvA experiments, leading to additional dHvA frequencies such as ($F_1+F_2$)/2. The absence of such MB effect indeed corroborates that at zero field, the intersections between two orbits [green dots in Fig.\,\ref{fig4}(d)] are Dirac points characterized by topological degeneracies implicitly guaranteed by the nonsymmorphic structure; the electron wavefunctions at the protected Dirac points are orthogonal, preventing the occurrence of either anticrossing or MB \cite{CoSi,MnSi}. As the gap opened by $H$ along $c$ is perturbatively small, the bands can retain their topological characters near the gap ( i.e., protected by the ``quasi-symmetry'' that preserves in the lower-order expansion of Hamiltonian \cite{Quasi-symmetry}). In such case, the MB process across the Dirac gap is still prohibited and only the original orbits $F_1$ and $F_2$ are resolvable, fully consistent with our observation.

  Our dHvA analysis strengthens the conclusion made on ARPES results \cite{LaNiGa2ARPES1,LaNiGa2ARPES2} and unambiguously confirms that the TRSB superconductor $\mathrm{LaNiGa_{\text{2}}}$ crystallize in a nonsymmorphic $Cmcm$ structure. Subsequently, our findings provide supporting evidence for the putative symmetry-enforced interband coupling \cite{QuanTopology} in $\mathrm{LaNiGa_{\text{2}}}$, which may lead to an exotic INT pairing state with full superconducting gap \cite{LaNiGa2TwoGap,LaNiGa2SCTheory}. Such enhancement of interband coupling is also revealed by recent transverse-field (TF) $\mu$SR data \cite{SundarMuon}. All these experimental evidences strongly imply that the INT pairing in $\mathrm{LaNiGa_{\text{2}}}$ arises from the topologically protected band degeneracies and the resulted reinforced interband electron transition, instead of strong electron correlation or magnetic fluctuations --- these are all absent herein \cite{LaNiGa2NMR} (to be noted, the Ni atoms in $\mathrm{LaNiGa_{\text{2}}}$ are in nominal $3d^{10}$ configuration \cite{SinghDFT} and the 3$d$ density of states at $E_F$ is relatively small, mostly attributing to the hybridization with Ga $p$ states \cite{SinghDFT,HaseLDA}); the pairing mechanism for such an unusual superconducting state is likely to be conventional electron-phonon coupling \cite{PhononCalc}. In this sense, $\mathrm{LaNiGa_{\text{2}}}$ probably serves as a unique representative of ``pure topological" TRSB superconductor. Meanwhile, there are still unanswered questions about this material, e.g., the superconductivity is relatively robust against electron irradiation \cite{LaNiGa2irradiation}, a typical behavior of sign-preserving singlet state but is unexpected for triplet pairing. Further studies involving more rigorous examination of the TRS in the superconducting state as well as tracking its evolution upon fine tuning (chemical substitution, pressure, uniaxial strain, etc) may help to nail down the nature of superconductivity in $\mathrm{LaNiGa_{\text{2}}}$.

\section{Conclusion}
	
  In summary, we synthesized the high-quality single crystals of $\mathrm{LaNiGa_{\text{2}}}$ by self-flux method. The detailed electronic structure underlying the unconventional TRSB superconductivity in this compound was probed by dHvA measurements in strong magnetic fields up to 33\,T. The resolved extremal FS orbits show good consistency with the FS model constructed by DFT calculations for the nonsymmorphic $Cmcm$ lattice symmetry. Moreover, a pair of loop orbits connected by topologically-protected Dirac nodes residing on the proposed nodal surface $k_z = \pi/c$ has been clearly resolved, underscoring the relevance of (nonsymmorphic) symmetry-enforced band degeneracy at $E_F$ in $\mathrm{LaNiGa_{\text{2}}}$, which is potentially essential for introducing the interband pairing potential required in the INT pairing scenario. Our results not only point towards an intimate link between the underlying space group symmetry and the topologically nontrivial electronic states in $\mathrm{LaNiGa_{\text{2}}}$, but also establish a useful method that can be applied to the exploration of novel superconductivity in a wide range of nonsymmorphic materials.
	
\section*{Acknowledgements}
  We thank Yang Gao for helpful discussions. This work was supported by the National Natural Science Foundation of China (Grant Nos. 12274390, 12488201, 12204449 and 12122411), the National Key R\&D Program of the MOST of China (Grant No. 2022YFA1602602), the Innovation Program for Quantum Science and Technology (2021ZD0302802), the Basic Research Program of the Chinese Academy of Sciences Based on Major Scientific Infrastructures (No. JZHKYPT-2021-08) and the Fundamental Research Funds for the Central Universities (Grant No. WK3510000014).


\begin{thebibliography}{99}

\bibitem{BCS} J. Bardeen, L. N. Cooper, and J. R. Schrieffer, Phys. Rev. 108, 1175 (1957).

\bibitem{AndersonGauge} P. W. Anderson, Phys. Rev. 130, 439 (1963).

\bibitem{AnnettSym} J. F. Annett, Adv. Phys. 39, 83 (1990).

%\bibitem {global} J. van Wezel and J. van den Brink, Phys. Rev. B 77, 064523 (2008).

\bibitem{UNSC} M. Sigrist and K. Ueda, Rev. Mod. Phys. 63, 239 (1991).

\bibitem{RecentTRSB} S. K. Ghosh, M. Smidman, T. Shang, J. F. Annett, A. D. Hillier, J. Quintanilla, and H. Yuan, J. Phys.: Condens. Matter 33, 033001 (2021).

%\bibitem {Cuprate} C. C. Tsuei and J. R. Kirtley, Rev. Mod. Phys. 72, 969 (2000).

%\bibitem {UPt3} R. Joynt and L. Taillefer, Rev. Mod. Phys. 74, 235 (2002).

\bibitem{CaPtAs} T. Shang, M. Smidman, A. Wang, L. J. Chang, C. Baines, M. K. Lee, Z. Y. Nie, G. M. Pang, W. Xie, W. B. Jiang, M. Shi, M. Medarde, T. Shiroka, and H. Q. Yuan, Phys. Rev. Lett. 124, 207001 (2020).

%\bibitem {SCSOC} M. Smidman, M. B. Salamon, H. Q. Yuan, and D. F. Agterberg, Rep. Prog. Phys. 80, 036501 (2017).

\bibitem{mSR1} S. J. Blundell, Contemp. Phys. 40, 175 (1999).

\bibitem{mSR2} J. E. Sonier, J. H. Brewer, and R. F. Kiefl, Rev. Mod. Phys. 72, 769 (2000).

\bibitem{kerr1} A. Kapitulnik, J. Xia, E. Schemm, and A. Palevski, New J. Phys. 11, 055060 (2009).

\bibitem{kerr2} E. R. Schemm, W. J. Gannon, C. M. Wishne, W. P. Halperin, and A. Kapitulnik, Science 345, 190 (2014).

\bibitem{ConvenTRSB} D. F. Agterberg, V. Barzykin, and L. P. $\mathrm{Gor'kov}$, Phys. Rev. B 60, 14868 (1999).

\bibitem{Tanaka} Y. Tanaka, J. Phys. Soc. Jpn. 70, 2844-2847 (2001).

\bibitem{Threeband} V. Stanev and Z. $\mathrm{Te\breve{s}anovi\acute{c}}$, Phys. Rev. B 81, 134522 (2010).

\bibitem{LeggettMode} S.-Z. Lin and X. Hu, Phys. Rev. Lett. 108, 177005 (2012).

%\bibitem {LaNiC2} A. D. Hillier, J. Quintanilla, and R. Cywinski, Phys. Rev. Lett. 102, 117007 (2009).

%\bibitem {LaNiC2SC} J. Quintanilla, A. D. Hillier, J. F. Annett, and R. Cywinski, Phys. Rev. B 82, 174511 (2010).

\bibitem{DaiLaFeAsOF} X. Dai, Z. Fang, Y. Zhou, and F.-C. Zhang, Phys. Rev. Lett. 101, 057008 (2008).

\bibitem{LaNiGa2TwoGap} Z. F. Weng, J. L. Zhang, M. Smidman, T. Shang, J. Quintanilla, J. F. Annett, M. Nicklas, G. M. Pang, L. Jiao, W. B. Jiang, Y. Chen, F. Steglich, and H. Q. Yuan, Phys. Rev. Lett. 117, 027001 (2016).

\bibitem{Interband1} T. Shang, S. K. Ghosh, M. Smidman, D. J. Gawryluk, C. Baines, A. Wang, W. Xie, Y. Chen, M. O. Ajeesh, M. Nicklas, E. Pomjakushina, M. Medarde, M. Shi, J. F. Annett, H. Yuan, J. Quintanilla, and T. Shiroka, npj Quantum Mater. 7, 35 (2022).

\bibitem{Interband2} S. K. Ghosh, P. K. Biswas, C. Xu, B. Li, J. Z. Zhao, A. D. Hillier, and X. Xu, Phys. Rev. Res. 4, L012031 (2022).

%\bibitem {ferminon} A. M. Black-Schaffer and C. Honerkamp, J. Phys. Condens. Matter 26, 423201 (2014).

%\bibitem {SrPtAs} H. Ueki, R. Tamura, and J. Goryo, Phys. Rev. B 99, 144510 (2019).

\bibitem{Topology1} N. P. Armitage, E. J. Mele, and A. Vishwanath, Rev. Mod. Phys. 90,015001 (2018).

\bibitem{Topology2} B. Q. Lv, T. Qian, and H. Ding, Rev. Mod. Phys. 93,025002 (2021).

\bibitem{QuanTopology} Y. Quan, V. Taufour, and W. E. Pickett, Phys. Rev. B 105, 064517 (2022).

\bibitem{LaNiGa2ARPES1} J. R. Badger, Y. Quan, M. C. Staab, S. Sumita, A. Rossi, K. P. Devlin, K. Neubauer, D. S. Shulman, J. C. Fettinger, P. Klavins, S. M. Kauzlarich, D. Aoki, I. M. Vishik, W. E. Pickett, and V. Taufour, Commun. Phys. 5, 22 (2022).

\bibitem{LaNiGa2ARPES2} M. Staab, R. Prater, S. Sreedhar, J. Byland, E. Mann, D. Zackaria, Y. Shi, H. J. Bowman, A. L. Stephens, M. C. Jung, A. S. Botana, W.  E. Pickett, V. Taufour, and I. Vishik,  Phys. Rev. B 110, 165115 (2024).

\bibitem{LaNiGa2NMR} P. Sherpa, I. Vinograd, Y. Shi, S. A. Sreedhar, C. Chaffey, T. Kissikov, M. C. Jung, A. S. Bontana, A. P. Dioguardi, R. Yamamoto, M. Hirata, G. Conti, S. Nemsak, J. R. Badger, P. Klavins, I. Vishik, V. Taufour, and N. J. Curro, Phys. Rev. B 109, 125113 (2024).

\bibitem{LaNiGa2SC} N. L. Zeng and W. H. Lee, Phys. Rev. B 66, 092503 (2002).

\bibitem{SundarMuon} S. Sundar, M. Yakovlev, N. Azari, M. Abedi, D. M. Broun, H. U. {\"O}zdemir, S. R. Dunsiger, D. Zackaria, H. Bowman, P. Klavins, Y. Shi, V. Taufour, and J. E. Sonier, Phys. Rev. B 109, 104517 (2023).

\bibitem{LaNiGa2Nonunitary} A. D. Hillier, J. Quintanilla, B. Mazidian, J. F. Annett, and R. Cywinski, Phys. Rev. Lett. 109, 097001 (2012).

\bibitem{LaNiGa2SCTheory} S. K. Ghosh, G. Csire, P. Whittlesea, J. F. Annett, M. Gradhand, B. $\mathrm{\acute{U}}$jfalussy, and J. Quintanilla, Phys. Rev. B 101, 100506(R) (2020).


\bibitem{LaNiGa2Cmmm} V. Romaka, Y. Grin, Y. Yarmolyuk, R. Skolozdra, and A. Jartys, Ukrains$'$kii Fizichnii Zhurnal 28, 227 (1983).

%\bibitem {TaNiTe5} Z. Hao, W. Chen, Y. Wang, J. Li, X. M. Ma, Y. J. Hao, R. Lu, Z. Shen, Z. Jiang, W. Liu, Q. Jiang, Y. Yang, X. Lei, L. Wang, Y. Fu, L. Zhou, L. Huang, Z. Liu, M. Ye, and D. Shen et al., Phys. Rev. B 104, 115158 (2021).

%\bibitem {TaPtTe5} S. Xiao, W. H. Jiao, Y. Lin, Q. Jiang, X. Yang, Y. He, Z. Jiang, Y. Yang, Z. Liu, M. Ye, D. Shen, and S. He, Phys. Rev. B 105, 195145 (2022).

\bibitem{Structure1} E. S. Makarov, and V. N. Bykov, Sov. Phys. Crystallogr. 4, 164 (1959).

\bibitem{Structure2} K. Oikawa, T. Kamiyama, H. Asano, Y. Onuki, and M. Kohgi, J. Phys. Soc. Jpn. 65, 3229 (1996).

%\bibitem{QO} H. Kim, K. F.  Wang, Y. Nakajima, R. W. Hu, S. Ziemak, P. Syers, L. M. Wang, H. Hodovanets, J. D. Denlinger, P. M. R. Brydon, D. F. Agterberg, M. A. Tanatar, R. Prozorov, and J. Paglione, Sci. Adv. 4, eaao4513 (2018).

%\bibitem {FeAsQO} S. E. Sebastian, J. Gillett, N. Harrison, P. H. C. Lau, D. J. Singh, C. H. Mielke, and G. G. Lonzarich, J. Phys. Condens. Matter 20, 422203 (2008).

\bibitem{DFT} G. Kresse and J. Furthm{\"u}ller, Phys. Rev. B 54, 11169 (1996).

\bibitem{PBE} J. P. Perdew, K. Burke, and M. Ernzerhof, Phys. Rev. Lett. 77, 3865 (1996).

\bibitem{Wannier} A. A. Mostofi, J. R. Yates, Y. S. Lee, I. Souza, D. Vanderbilt, and N. Marzari, Comput. Phys. Commun. 178, 685 (2008).

\bibitem{SKEAF} P. M. C. Rourke and S. R. Julian, Comput. Phys. Commun. 183, 324 (2012).

\bibitem{Shoenberg} D. Shoenberg, \textit{Magnetic Oscillations in Metals}, (Cambridge University Press, Cambridge, England, 1984)

\bibitem{FeGe} Kaixin Tang, Hanjing Zhou, Houpu Li, Senyang Pan, Xueliang Wu, Hongyu Li, Nan Zhang, Chuanying Xi, Jinglei Zhang, Aifeng Wang, Xiangang Wan, Ziji Xiang, and Xianhui Chen, Phys. Rev. Research 6, 013276 (2024).

\bibitem{breakdown} M. H. Cohen and L. M. Falicov, Magnetic Breakdown in Crystals, Phys. Rev. Lett. 7, 231 (1961).

%\bibitem{breakdown2} A. Alexandradinata and L. Glazman, Phys. Rev. B 97, 144422 (2018).

%\bibitem{breakdown} C. S. M{\"u}ller, T. Khouri, M. R. van Delft, S. Pezzini, Y. T. Hsu, J. Ayres, M. Breitkreiz, L. M. Schoop, A. Carrington, N. E. Hussey, and S. Wiedmann, Phys. Rev. Res. 2, 023217 (2020).

\bibitem{SinghDFT} D. Singh, Phys. Rev. B 86, 174507 (2012).

\bibitem{HaseLDA} I. Hase and T. Yanagisawa,  J. Phys. Soc. Jpn. 81, 103704 (2012).

\bibitem{Nodal_surface} Q.-F. Liang, J. Zhou, R. Yu, Z. Wang, and H. Weng, Phys. Rev. B 93, 085427 (2016).

\bibitem{Chambers} R. G. Chambers, Proc. Phys. Soc. London 88, 701 (1966).

\bibitem{Quasi-symmetry} C. Guo, L. Hu, C. Putzke, J. Diaz, X. Huang, K. Manna, F.-R. Fan, C. Shekhar, Y. Sun, C. Felser, C. Liu, B. A. Bernevig, and P. J. W. Moll, Nat. Phys. 18, 813 (2022).

\bibitem{Tunneling} E. I. Blount, Phys. Rev. 126, 1636 (1962).

\bibitem{MnSi} M. A. Wilde, M. Dodenh{\"o}ft, A. Niedermayr, A. Bauer, M. M. Hirschmann, K. Alpin, A. P. Schnyder, and C. Pfleiderer, Nature (London) 594, 374 (2021).

\bibitem{CoSi} N. Huber, K. Alpin, G. L. Causer, L. Worch, A. Bauer, G. Benka, M. M. Hirschmann, A. P. Schnyder, C. Pfleiderer, and M. A. Wilde, Phys. Rev. Lett. 129, 026401 (2022).

\bibitem{PhononCalc} H. M. T{\"u}t{\"u}nc{\"u} and G. P. Srivastava, Appl. Phys. Lett. 104, 022603 (2014).

\bibitem{LaNiGa2irradiation} S. Ghimire, K. R. Joshi, E. H. Krenkel, M. A. Tanatar, Yunshu Shi, M. Ko$\mathrm{\acute{n}}$czykowski, R. Grasset, V. Taufour, P. P. Orth, M. S. Scheurer, and R. Prozorov, Phys. Rev. B 109, 024515 (2024).

%\bibitem {LaNiGa2Cmmmelec} I. Hase and T. Yanagisawa, J. Phys. Soc. Jpn. 81, 103704 (2012).

\end{thebibliography}
\end{document}